\documentclass[aps,prb,twocolumn,showpacs]{revtex4}
\usepackage{amssymb}
\usepackage{graphicx}

\begin{document}

\title{\textbf{\ Thermal transport in one-dimensional spin heterostructures}}
\author{Liliana Arrachea$^1$ , Gustavo S. Lozano$^1$ and A. A. Aligia$^2$}

\address{$^1$ Departamento de F\'{\i}sica, FCEyN, Universidad de Buenos Aires, 
Pabell\'on 1,Ciudad Universitaria, 1428, Buenos Aires, Argentina.} 
\address{$^2$ Comisi\'{o}n Nacional de Energ{\'{\i }}a At\'{o}mica, 
Centro At\'omico Bariloche and Instituto Balseiro,  8400 S.C. de Bariloche, Argentina.}

\begin{abstract}
We study heat transport in a one-dimensional inhomogeneous quantum spin $1/2$ system.
It consists of a 
finite-size XX 
spin chain coupled at its ends to  
semi-infinite XX and XY chains at different temperatures, which play the
role of heat and spin reservoirs.
After using the Jordan-Wigner transformation we map the original spin Hamiltonian
into a fermionic Hamiltonian, which contains normal and pairing terms. We find
the expressions for the heat currents and solve the problem with a non-equilibrium
Green's function formalism. We analyze the behavior of the heat currents as functions
of the model parameters. When finite magnetic fields are applied at the two 
reservoirs, the system exhibits rectifying effects in the heat flow. 
\end{abstract}

\maketitle

\section{ Introduction}

In the last decade there has been a renewed interest related to the research
of thermal transport in one-dimensional magnetic systems \cite{Solo,dash} . Most
of these studies have been motivated by unusual high values of thermal
conductance of some materials, as for example reported in Ref. 
\onlinecite{lista2}. From the theoretical side, several works calculating
and discussing thermal conductivity in different model Hamiltonians have
also appeared \cite{lista5}, \cite{lista3}. On the other hand, spin Hamiltonians
provide the natural scenario to implement quantum computation. This motivated 
interesting proposals of a variety of physical systems, like arrays of quantum
dots, optical lattices and nuclear magnetic resonace experiments which are architectured to effectively
behave like one-dimensional spin systems. \cite{design}

Most of  the theoretical studies on heat transport in spin systems are performed within 
the linear response regime,
assuming a very small temperature gradient $\bigtriangledown T$, and using a
Kubo formula. Altough this formula is widely used in calculating thermal transport properties it has several conceptual difficulties, particularly
if $\bigtriangledown T$ is not small \cite{mahan,dash}. The thermal averaging
assumes a constant temperature, and (as in all Kubo formulas), the time
dependence is governed by the total Hamiltonian $H=H_{0}+H'$, including the perturbation 
$H' \sim \bigtriangledown T$ , while the thermodynamic averaging is done with 
$H_{0}$, assuming a fast evolution. For finite $\bigtriangledown T$ the
separation of $H$ into an unperturbed part $H_{0}$ and a perturbation $H'$ 
is in principle ambiguous.
On the other hand, real experiments are performed by coupling the system under study to
macroscopic systems with  well defined temperatures that act as reservoirs. 
 Therefore, it seems desirable to develop
alternative approaches designed to treat systems out of equilibrium. 
Important progress has been achieved by studying the properties of non-equilibrium steady states of XX and XY chains within an algebraic 
setting which allows one to obtain explicit analytical expressions for different quantities such as entropy production \cite{pillet}.
Recently, the usefulness of Kubo formula for the investigation of heat transport in quantum systems has been discussed \cite{gemmer1}
and 
the investigation of energy transport through spin systems beyond Kubo formula has been 
addressed on the basis of master equations and quantum Monte Carlo methods. \cite{gemmer2}
Nevertheless, it is not clear how to extend these ideas to more general situations.

In this work we address the problem of thermal transport in one dimensional magnetic systems from a different perspective.
We study 
a problem that to the best of our knowledge has not
been considered yet. We study the heat flow through a \textit{spin chain
heterostructure}, which we generically define as a set of finite or
semi-infinite spin chains attached by their ends. 
Each piece of the heterostructure
can be in principle described by a {\it different} Hamiltonian.
For definiteness, we shall
consider in this work a finite central system connected to two (left and
right) semi-infinite chains, with a\textit{\ finite} temperature difference
between them\cite{karevski,gemmer2}. 
This problem is the thermal analog of electronic transport
through mesoscopic structures, nanodevices or molecules connected to conducting 
leads with a finite
applied bias voltage, a subject of intense research in recent years. In
fact, this type of set-up is the common situation 
found in 
the study of
charge transport in electronic systems, where a central system is connected
to charge reservoirs that also act as thermal baths. This is the basis of
the "Landauer" approach, which is one of the most common frameworks to study
transport properties of meso- and nano- devices in the last years.
\cite{transporte}

In addition, as it is well known, one-dimensional spin $1/2$ systems can be
mapped, via the Jordan-Wigner transformation to fermionic systems. Thus,
the model under investigation is equivalent to an electronic heterostructure
where very well established techniques, as the Schwinger-Keldysh non
equilibrium Green's functions method can be applied \cite{Kel}.

We focus on 
a simple device where the central and right parts of
the system can be described by XX spin 1/2 chains, and the left part corresponds
to an anisotropic XY chain. The Jordan-Wigner transformation maps these
models into bilinear fermionic systems, rendering the
theoretical study simpler. We show that this simple device presents an
interesting physical effect: due to a mechanism reminiscent of Andreev
reflections in superconductors, this device could act as thermal diode.
This rectifying effect might be useful for applications.
We argue that this is a generic feature, which remains valid for 
anisotropic XYZ models.

The paper is organized as follows: in Section II we present the model, and
the non-equilibrium formalism based on Green's functions. 
Section III contains the numerical results. Section IV is a
summary and discussion.

\section{The formalism}

\subsection{Model}

\begin{figure}
\centering   
\includegraphics
[width=80mm,height=50mm,angle=0]{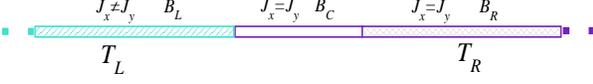} {\small {} }
\caption{{
(Color online) Schematic digram showing the set up of the system.
The right and central chains correspond to $XX$ models while the left chain is described by an $XY$ model. 
Left and right chains are taken semi-infinite at temperatures $T_L$ and $T_R$ respectively.}}
\label{dibujo}
\end{figure}

We consider a system of three one-dimensional spin 1/2 chains coupled through
their ends as shown in Fig. \ref{dibujo}. The system is described by the following Hamiltonian $%
H=H_{L}+H_{C}+H_{R}+H_{coup}$, where 
\begin{eqnarray}
H_{\alpha } &=&J_{\alpha }\sum_{i=1}^{N_{\alpha }-1}[S_{\alpha
,i}^{x}S_{\alpha ,i+1}^{x}+\lambda _{\alpha }S_{\alpha ,i}^{y}S_{\alpha
,i+1}^{y}]  \nonumber \\
&&-B_{\alpha }\sum_{i=1}^{N_{\alpha }}S_{\alpha ,i}^{z},  \nonumber \\
H_{coup} &=&J^{\prime }[S_{L,1}^{x}S_{C,1}^{x}+S_{L,1}^{y}S_{C,1}^{y}]+ 
\nonumber \\
&&J^{\prime }[S_{R,1}^{x}S_{C,N_{C}}^{x}+S_{R,1}^{y}S_{N_{C},1}^{y}],
\label{ham1}
\end{eqnarray}%
where in the first line, the index $\alpha =L,C,R$ (left, center, right)
labels the different chains. Each chain is characterized by its
nearest-neighbor exchange $J_{\alpha }$, with anisotropy $\lambda _{\alpha }$
for the ratio of the interaction along the $y$ direction with respect to
that along $x,$ and the magnetic field $B_{\alpha }$ applied along the $z$
direction. The isotropic case for a given chain corresponds to $\lambda
_{\alpha }=1$. The Hamiltonian $H_{coup}$ describes the coupling between the
central chain, and the left and right ones. In terms of the representation
for the spin operators: $S^{\pm }=S^{x}\pm iS^{y}$, the above Hamiltonian
reads 
\begin{eqnarray}
H_{\alpha } &=&\frac{J_{\alpha }}{4}\sum_{l=1}^{N_{\alpha }-1}[(1-\lambda
_{\alpha })(S_{\alpha ,l}^{+}S_{\alpha ,l+1}^{+}+S_{\alpha ,l}^{-}S_{\alpha
,l+1}^{-})  \nonumber \\
&&+(1+\lambda _{\alpha })(S_{\alpha ,l}^{+}S_{\alpha ,l+1}^{-}+S_{\alpha
,l}^{-}S_{\alpha ,l+1}^{+})]  \nonumber \\
&&-B_{\alpha }\sum_{l=1}^{N_{\alpha }}S_{\alpha ,l}^{z},  \nonumber \\
H_{coup} &=&\frac{J^{\prime }}{2}%
[S_{L,1}^{+}S_{C,1}^{-}+S_{L,1}^{-}S_{C,1}^{+}]+  \nonumber \\
&&\frac{J^{\prime }}{2}%
[S_{R,1}^{-}S_{C,N_{C}}^{+}+S_{R,1}^{+}S_{N_{C},1}^{-}].  \label{ham2}
\end{eqnarray}%
For the isotropic case, $\lambda _{\alpha }=1$ only the flip-flop terms with
products of one raising and and one lowering spin operators survive.

\bigskip

We now introduce the Jordan-Wigner transformation to map the spin 1/2
Hamiltonian into a fermionic Hamiltonian through: $S_{\alpha
,l}^{+}=f_{\alpha ,l}^{\dagger }\exp (i \pi \sum_{j}^{\prime }f_{\alpha
,j}^{\dagger }f_{\alpha ,j})$, where $\sum_{j}^{\prime }$ denotes a sum over
all the positions located at the left of the position $\alpha ,l$.
Similarly, the other spin operators transform as $S_{\alpha ,l}^{-}=\exp
(-i \pi \sum_{j}^{\prime }f_{\alpha ,j}^{\dagger }f_{\alpha ,j})f_{\alpha ,l}$
and $S_{\alpha ,l}^{z}=f_{\alpha ,l}^{\dagger }f_{\alpha ,l}-1/2$, where the
operators $f_{l},f_{l}^{\dagger }$ obey fermionic commutation rules: $%
\{f_{\alpha ,l},f_{\alpha ^{\prime },l^{\prime }}^{\dagger }\}=\delta
_{l,l^{\prime }}\delta _{\alpha ,\alpha ^{\prime }}$, and $\{f_{\alpha
,l}^{\dagger },f_{\alpha ^{\prime },l^{\prime }}^{\dagger }\}=\{f_{\alpha
,l},f_{\alpha ^{\prime },l^{\prime }}\}=0$. Substituting in the Hamiltonian (%
\ref{ham2}), we get 
\begin{eqnarray}
H_{\alpha } &=&\sum_{l=1}^{N_{\alpha }-1}\{w_{\alpha }[f_{\alpha
,l}^{\dagger }f_{\alpha ,l+1}+f_{\alpha ,l+1}^{\dagger }f_{\alpha ,l}] 
\nonumber \\
&&+\Delta _{\alpha }[f_{\alpha ,l}^{\dagger }f_{\alpha ,l+1}^{\dagger
}+f_{\alpha ,l+1}f_{\alpha ,l}]\}  \nonumber \\
&&-\mu _{\alpha }\sum_{l=1}^{N_{\alpha }}f_{\alpha ,l}^{\dagger }f_{\alpha
,l},  \nonumber \\
H_{coup} &=&w^{\prime }[f_{L,1}^{\dagger }f_{C,1}+f_{C,1}^{\dagger }f_{L,1}]+
\nonumber \\
&&w^{\prime }[f_{C,N_{C}}^{\dagger }f_{R,1}+f_{R,1}^{\dagger }f_{C,N_{C}}],
\label{ham3}
\end{eqnarray}%
where $w_{\alpha }=J_{\alpha }(1+\lambda _{\alpha })/4$, $\Delta _{\alpha
}=J_{\alpha }(1-\lambda _{\alpha })/4$, $\mu _{\alpha }=B_{\alpha }^{z}$,
and $w^{\prime }=J^{\prime }/2$. Therefore, in the language of fermionic operators,
the Hamiltonian contains \textquotedblleft normal\textquotedblright
terms,
with a creation and a destruction operator, 
as well as  \textquotedblleft anomalous\textquotedblright\
terms, with two creation 
or 
two destruction operators.
The normal ones are a hopping term between nearest neighbors ($w_{\alpha}$), which is
originated in the flip-flop spin terms and a chemical potential ($\mu_{\alpha}$)
coupled to the fermionic density,
which comes from the magnetic field pointing along the $z$-direction. The anomalous terms
($\Delta_{\alpha}$) are 
similar to those of a
one-dimensional Hamiltonian with a gap function with $p$-wave symmetry,
decoupled in the Bardeen-Cooper-Schrieffer (BCS) approximation and
are 
originated by 
the anisotropy between the $X$ and $Y$ exchange interaction.

Inspired in this analogy, we 
focus our study on 
a junction between a chain
with isotropic interactions ($XX$ spin chain) and an anisotropic one ($XY$
spin chain), which in the fermionic language is similar to a
normal-superconductor junction. Such a situation is realized in a
configuration with $\lambda _{R}=\lambda _{C}=0$ and $\lambda _{L}\neq 0$.
We also assume that the left and right chains are at temperatures 
$T_{L}$ and $T_{R}$ respectively and they are both of infinite length( $%
N_{L}\rightarrow \infty $ and $N_{R}\rightarrow \infty $)

\subsection{Energy balance}

A consistent procedure to define an expression 
for the heat current from first principles, is to  analyze the
evolution of the energy stored in a small volume of the system
and derive the corresponding 
equation for the 
conservation 
of the energy. \cite{liliheat}
For the present Hamiltonian we choose an elementary volume  containing two
nearest-neighbor positions of the chain. We place the volume
enclosing the sites $l,l+1$ within the central (XX) chain, which in the
fermionic language contains only normal terms.  
We work in units where $\hbar=1$. 
The equation 
for the conservation of 
the energy enclosed by this volume is 
\begin{eqnarray}
\frac{dE_{l,l+1}}{dt} &=&\frac{J_{C}}{2}\frac{d}{dt}\langle
S_{l}^{+}S_{l+1}^{-}+S_{l}^{-}S_{l+1}^{+}\rangle -B_{C}^{z}\frac{d\langle
S_{l}^{z}\rangle }{dt}  \nonumber \\
&=&w_{C}\frac{d}{dt}\langle f_{C,l}^{\dagger }f_{C,l+1}+f_{C,l+1}^{\dagger
}f_{C,l}\rangle -
\nonumber \\
&&B_{C}^{z}\frac{d}{dt}\langle f_{C,l}^{\dagger
}f_{C,l}\rangle   \nonumber \\
&=&-iw_{C}\langle \lbrack H,f_{C,l}^{\dagger }f_{C,l+1}+f_{C,l+1}^{\dagger
}f_{C,l}]\rangle   \nonumber \\
&&+i\mu _{C}\langle \lbrack H,f_{C,l}^{\dagger
}f_{C,l}]\rangle=J^Q_{l+1,l+2}-J^Q_{l-1,l},  \label{de}
\end{eqnarray}%
where $J^Q_{l,l+1}$ is the heat current flowing from $l$ to $l+1$, which
in the present setup coincides with the energy current. Its explicit
expression is obtained from the evaluation of the above commutator, which
gives 
\begin{eqnarray}
J^Q &=&J^Q_{l,l+1}=J^Q_{l-1,l}  \nonumber \\
&=&i\varepsilon _{l,l+1}^{C}(\varepsilon _{l-1,l}^{C}\langle
f_{C,l-1}^{\dagger }f_{C,l+1}-f_{C,l+1}^{\dagger }f_{C,l-1}\rangle  
\nonumber \\
&&+\varepsilon _{l+1,l+1}^{C}\langle f_{C,l+2}^{\dagger
}f_{C,l+1}-f_{C,l+1}^{\dagger }f_{C,l+2}\rangle ),  \label{j1}
\end{eqnarray}%
where $\varepsilon _{l,l^{\prime }}^{C}$ denotes the matrix element $%
l,l^{\prime }$ of the Hamiltonian $H_{C}$. In order to evaluate the above
current it is convenient to introduce the lesser Green's functions
\begin{equation}
G_{\alpha l,\beta l^{\prime }}^{<}(t,t^{\prime })=i\langle f_{\beta
l^{\prime }}^{\dagger }(t^{\prime })f_{\alpha l}(t)\rangle ,  \label{gle}
\end{equation}%
thus
\begin{eqnarray}
J^Q &=&2\mbox{Re}\{\varepsilon
_{l,l+1}^{C}G_{Cl+1,Cl-1}^{<}(t,t)\varepsilon _{l-1,l}^{C}  \nonumber \\
&&+\varepsilon _{l+1,l+1}^{C}G_{Cl+1,Cl+2}^{<}(t,t)\varepsilon
_{l+2,l+1}^{C}\}.  \label{je}
\end{eqnarray}%
The lesser Green's functions are one of the basic elements within Keldysh
non-equilibrium Green's function formalism \cite{Kel}. They are evaluated by
solving  the equations of motion (Dyson's equations), which
for our model can be written as follows 
\begin{equation}
\sum_{k}G_{Cl,Ck}^{<}(\omega )[\delta _{k,l^{\prime }}\omega -\varepsilon
_{k,l^{\prime }}^{C}]=0  \label{gle2}
\end{equation}%
for coordinates $l,l^{\prime }$ lying within the central chain. We have used
the stationary property of the system, as a consequence of which the Green's
functions depend on the difference $t-t^{\prime }$, which allows us to
transform: $G_{j,j^{\prime }}^{<}(t-t^{\prime })=\int_{-\infty }^{+\infty
}d\omega /(2\pi )e^{-i\omega (t-t^{\prime })}G_{j,j^{\prime }}^{<}(\omega )$.
Thus, using the above equation in Eq. (\ref{je}) the energy current can be
also expressed in the following way 
\begin{eqnarray}
J^Q & = & 2\mbox{Re}\{\int_{-\infty }^{+\infty }\frac{d\omega }{2\pi }\omega
[\varepsilon _{l,l+1}^{C}G_{Cl+1,Cl}^{<}(\omega ) \nonumber \\
& & +\varepsilon _{l+1,l+1}^{C}G_{Cl+1,Cl+1}^{<}(\omega ) ]\}.  \label{jqq1}
\end{eqnarray}
However $\mbox{Re}
\{G_{j+1,j+1}^{<}(\omega )\}=0$. Thus, the heat current reduces to  
\begin{equation}
J^Q=2\mbox{Re}\{\int_{-\infty }^{+\infty }\frac{d\omega }{2\pi }\omega
\varepsilon _{l,l+1}^{C}G_{Cl+1,Cl}^{<}(\omega ) \}.  \label{jqq}
\end{equation}%
 Similarly, if we evaluate the heat current
through the contacts $L-C$ and $C-R$, we find  
\begin{eqnarray}
J^Q &=&2w^{\prime }\mbox{Re}\{\int_{-\infty }^{+\infty }\frac{d\omega }{%
2\pi }\omega G_{L1,C1}^{<}(\omega )\},  \nonumber \\
&=&2w^{\prime }\mbox{Re}\{\int_{-\infty }^{+\infty }\frac{d\omega }{2\pi }%
\omega G_{CN,R1}^{<}(\omega )\}.  \label{jq2}
\end{eqnarray}

\subsection{Solving Dyson's equations}

In order to evaluate $G^{<}$ and the heat current we follow a treatment
close to that presented in Ref. \onlinecite{lilisup}. We define the retarded
\textquotedblleft normal\textquotedblright\ and \textquotedblleft
Gorkov\textquotedblright\ Green's functions 
\begin{eqnarray}
G_{\alpha l,\beta l^{\prime }}^{R}(t,t^{\prime }) &=&-i\Theta (t-t^{\prime
})\langle \{f_{\alpha ,l}(t),f_{\beta ,l^{\prime }}^{\dagger }(t^{\prime
})\}\rangle ,  \nonumber \\
F_{\alpha l,\beta l^{\prime }}^{R}(t,t^{\prime }) &=&-i\Theta (t-t^{\prime
})\langle \{f_{\alpha ,l}^{\dagger }(t),f_{\beta ,l^{\prime }}^{\dagger
}(t^{\prime })\}\rangle .  \label{gf}
\end{eqnarray}%

Before writing down the Dyson's equations satisfied the these 
\textquotedblleft full\textquotedblright Green's function let us define
the following
\textquotedblleft free\textquotedblright particle $\hat{g}_{\alpha }^{0}(\omega )$
and hole $\hat{\overline{g}}_{\alpha }^{0}(\omega )$ Green's functions
\begin{eqnarray}
&&[\hat{g}_{\alpha }^{0}(\omega )]_{\alpha l,\alpha l^{\prime }}^{-1}=\delta
_{l,l^{\prime }}(\omega + i \eta) -\varepsilon _{l,l^{\prime }}^{\alpha }  \nonumber \\
&&[\hat{\overline{g}}_{\alpha }^{0}(\omega )]_{\alpha l,\alpha l^{\prime
}}^{-1}=\delta _{l,l^{\prime }}(\omega+ i \eta) +\varepsilon _{l,l^{\prime }}^{\alpha
},  \label{g0}
\end{eqnarray}%
with $\eta= 0^+$.

From now on we will work in Fourier space and we will not write explicitly 
the $\omega$ dependence of Green's functions unless necessary.
For the left chain, we also define the functions $\hat{G}_{L}^{0}$, 
$\hat{\overline{G}}_{L}^{0}$ containing the paring term contribution, through the relations
\begin{eqnarray}
&&[\hat{G}_{L}^{0}]^{-1}=[g_{L}^{0}]^{-1}-\hat{\Delta}_{L}%
\hat{\overline{g}}_{L}^{0}\hat{\Delta}_{L},  \nonumber \\
&&[\hat{\overline{G}}_{L}^{0}]^{-1}=[\overline{g}_{L}^{0}]^{-1}-\hat{\Delta}_{L}
\hat{g}_{L}^{0}\hat{\Delta}_{L},  
\end{eqnarray}
We also introduce
\begin{eqnarray}
&& \hat{g}^{0}=\sum_{\alpha =L,C,R}\hat{g}_{\alpha }^{0} ,\;\;
\hat{\overline{g}}^{0}=\sum_{\alpha =L,C,R}\hat{\overline{g}}_{\alpha }^{0},\nonumber \\ 
&&\hat{G}^{0}=\hat{G}_{L}^{0}+\hat{g}_{C}^{0}+%
\hat{g}_{R}^{0},  \nonumber \\ &&
\hat{\overline{G}}^{0}=\hat{\overline{G}}_{L}^{0}+\hat{%
\overline{g}}_{C}^{0}+\hat{\overline{g}}_{R}^{0}, 
\nonumber \\
&&\hat{F}_{L}^{0}=\hat{\overline{G}}^{0}\hat{\Delta}_{L}%
\hat{g}_{L}^{0},  \nonumber \\
&&\hat{\overline{F}}_{L}^{0}=\hat{G}^{0}\hat{\Delta}_{L}%
\hat{\overline{g}}_{L}^{0}.
\end{eqnarray}%
Here $\varepsilon _{l,l^{\prime }}^{\alpha }$ and $\Delta _{\alpha l,\alpha l^{\prime
}}^{ }$ are matrices defined on the coordinates of the chain $\alpha
=L,C,$ or $R$, containing respectively, the normal and anomalous elements of
the Hamiltonian. In the case we are studying only $\hat{\Delta}_{L}$ is
non-vanishing.

To obtain these Green's functions, we write the following Dyson's equation
which relate them with Green's functions of the \textquotedblleft
disconnected" chains $\hat{g}_{\alpha }^{0}$ and $\hat{\overline{g}}%
_{\alpha }^{0}$ (see below) and the matrix elements of the contacts 
\begin{eqnarray}
&&\{[\hat{g}_{L}^{0}]^{-1}+[\hat{g}_{C}^{0}]^{-1}+[\hat{g}%
_{R}^{0}]^{-1}-  \nonumber \\ &&
[\hat{W}]\}\hat{G}^{R}-\hat{\Delta}_{L}\hat{F}%
^{R}=\hat{1},  \label{dyret1} \\
&&\{[\hat{\overline{g}}_{L}^{0}]^{-1}+[\hat{\overline{g}}%
_{C}^{0}]^{-1}+[\hat{\overline{g}}_{R}^{0}]^{-1}+ 
\nonumber \\
&&[\hat{W}]\}\hat{F}^{R}-\hat{\Delta}_{L}\hat{G}%
^{R}=\hat{0}.  \label{dyret2}
\end{eqnarray}%
 The matrix $\hat{W}=\hat{W}_{L}+\hat{W}_{R}$ contains the
matrix elements of $H_{cont}$ describing the connections between 
the central and left parts  and between the central and right parts. 

\bigskip 

The above equations can be rewritten in a more convenient form by recourse
to the following procedure. From Eq. (\ref{dyret2}) 
\begin{equation}
\hat{F}^{R}=\hat{\overline{g}}^{0}[\hat{\Delta}_{L}\hat{G}%
^{R}-\hat{W}\hat{F}^{R}],  \label{f}
\end{equation}%
\begin{equation}
\hat{G}^{R}=\hat{g}^{0}+\hat{g}^{0}[\hat{\Delta}%
_{L}\hat{F}^{R}+\hat{W}\hat{G}^{R}],  \label{g}
\end{equation}%

Substituting (\ref{f}) in (\ref{dyret1}) and (\ref{g}) in (\ref{dyret2}) one
obtains 
\begin{eqnarray}
& &\hat{G}^{R} =\hat{G}^{0}(1+\hat{W}\hat{%
G}^{R})+\hat{\overline{F}}_{L}^{0}\hat{W}\hat{F}^{R},  \label{dyret3} \\
& &\hat{F}^{R} =\hat{F}_{L}^{0}(1+%
\hat{W}\hat{G}^{R})-\hat{\overline{G}}^{0}\hat{W}\hat{F}%
^{R},  \label{dyret4}
\end{eqnarray}%

Let us now consider Eq. (\ref{dyret3}) for the following particular
coordinates  
\begin{eqnarray}
G_{Cl,Cl^{\prime }}^{R}&=&g_{C,l,l^{\prime }}^{0}+g_{C,l,1}^{0}w^{\prime }G_{L1,Cl^{\prime }}^{R} 
\nonumber \\
&&+g_{C,l,N}^{0}w^{\prime }G_{R1,Cl^{\prime }}^{R},
\label{con1} \\
G_{L1,Cl^{\prime }}^{R} &=&G_{L,1,1}^{0}w^{\prime
}G_{C1,Cl^{\prime }}^{R} \nonumber \\
&&+\overline{F}_{L,1,1}^{0}w^{\prime }F_{C1,Cl^{\prime
}}^{R},  \label{con2} \\
G_{R1,Cl^{\prime }}^{R} &=&g_{R,1,1}^{0}w^{\prime
}G_{CN,Cl^{\prime }}^{R}.  \label{con3}
\end{eqnarray}%
Substituting Eqs. (\ref{con2}) and (\ref{con3}) in Eq. (\ref{con1}) it is
easy to see that the Dyson's equation for the two indices corresponding to
coordinates of $C$ can be written as follows 
\begin{eqnarray}
&&\{[\hat{g}_{C}^{0}]^{-1}-\hat{\Sigma}^{R,gg}\}\hat{G}%
_{C}^{R}  
+\hat{\Sigma}^{R,gf}\hat{F}_{C}^{R}=\hat{1},
\label{dyret5} \\
&&\{[\hat{\overline{g}}_{L}^{0}]^{-1}-\hat{\Sigma}^{R,ff}\}%
\hat{F}_{C}^{R}  +\hat{\Sigma}^{R,fg}\hat{G}_{C}^{R}=\hat{0},
\label{dyret6}
\end{eqnarray}
where the matrices of the above equations have sizes $N_{C}\times N_{C}$ and
elements corresponding to the coordinates of the central chain. The
\textquotedblleft self-energy\textquotedblright\ matrices are  
\begin{eqnarray}
\Sigma _{l,l^{\prime }}^{R,ff} &=&\delta _{l,l^{\prime }}|w^{\prime
}|^{2}[\delta _{l,1}\overline{G}_{L,1,1}^{0}+\delta _{l,N_{C}}%
\overline{g}_{R,1,1}^{0}  \nonumber \\
\Sigma _{l,l^{\prime }}^{R,gg} &=&\delta _{l,l^{\prime }}|w^{\prime
}|^{2}[\delta _{l,1}G_{L,1,1}^{0}+\delta
_{l,N_{C}}g_{R,1,1}^{0}]  \nonumber \\
\Sigma _{l,l^{\prime }}^{R,gf} &=&\delta _{l,l^{\prime }}|w^{\prime
}|^{2}\delta _{l,1}\overline{F}_{L,1,1}^{0},  \nonumber \\
\Sigma _{l,l^{\prime }}^{R,fg} &=&\delta _{l,l^{\prime }}|w^{\prime
}|^{2}\delta _{l,1}F_{L,1,1}^{0}.  \label{sigmas}
\end{eqnarray}%
The explicit expressions for these functions imply the evaluation of all the
functions appearing in the right hand sides of (\ref{sigmas}). Notice that
these functions have been defined from manipulations of the Dyson's
equations corresponding to $H_{L}$ or $H_{R}$ \emph{isolated} from the
central chain. We indicate a procedure for the calculation of these
functions in appendix A. Note also that since the right and left parts of
the system are held at two different but constant temperatures, these Green's 
functions can be calculated at equilibrium.
The advantage of the above representation becomes clear by writing (\ref%
{dyret6}) as 
\begin{eqnarray}
&&\hat{F}_{C}^{R}=\hat{\overline{g}}_{C}\hat{\Sigma}%
^{R,fg}\hat{G}^{R},  \nonumber \\
&&[\hat{\overline{g}}_{C}]^{-1}=[\hat{\overline{g}}_{C}^{0}]^{-1}-\hat{\Sigma}^{R,ff},  \label{fc}
\end{eqnarray}%
and substituting it in (\ref{dyret5}). The result leads to the solution of
the retarded normal Green's function within $C$  
\begin{eqnarray}
\hat{G}_{C}^{R} &=&\{[\hat{g}_{C}^{0}]^{-1}-\hat{\Sigma}%
_{\mbox{{\it eff}}}^{-1},  \nonumber \\
\hat{\Sigma}^{R}_{\mbox{{\it eff}}} &=&\hat{\Sigma}^{R,gg}+\hat{\Sigma}%
^{R,gf}\hat{\overline{g}}_{C}\hat{\Sigma}^{R,fg}.
\label{gc}
\end{eqnarray}

The results obtained so far correspond to the retarded Green's functions and
self energies.  The lesser Green's function with coordinates within $C$ can
be easily obtained by recourse to Langreth rules \cite{langr,hern}. In
particular one obtains \cite{lilisup}%
\begin{equation}
\hat{G}_{C}^{<}=\hat{G}_{C}^{R}\hat{\Sigma}^{<}_{\mbox{{\it eff}}}\hat{G}_{C}^{A},  \label{gcles}
\end{equation}%
where the advanced Green's function is obtained from the retarded one, by
means of the relation $\hat{G}_{C}^{A}(\omega )=[\hat{G}_{C}^{R}(\omega
)]^{\dagger }$ and the lesser component of the \textquotedblleft
self-energy\textquotedblright\ is 
\begin{eqnarray}
& &\hat{\Sigma}^{<}_{\mbox{{\it eff}}} =\hat{\Sigma}^{<,gg}+  
\hat{\Lambda}^{R}\hat{\Sigma}^{<,fg}+
\nonumber \\ & &\hat{\Sigma}%
^{<,gf}\hat{\Lambda}^{A} +
\hat{\Lambda}^{R}\hat{\Sigma}^{<,ff}\hat{\Lambda}%
^{A},  \label{sigefles}
\end{eqnarray}%
with $\hat{\Lambda}^{R}(\omega )=\hat{\Sigma}_{\alpha }^{R,gf}(\omega )\hat{%
\overline{g}}_{C}(\omega )$ and $\hat{\Lambda}^{A}(\omega )=[\hat{\Lambda}%
^{R}(\omega )]^{\dagger }$. 
The self-energies have components
\[
 \Sigma
_{l,l^{\prime }}^{<,\nu ,\nu ^{\prime }}(\omega )=i\delta _{l,l^{\prime
}}[\delta _{l,1}\Gamma _{L}^{\nu ,\nu ^{\prime }}(\omega )f_{L}(\omega
)+\delta _{l,N}\Gamma _{R}^{\nu ,\nu ^{\prime }}(\omega )f_{R}(\omega )]
\nonumber
\]
being $\Gamma _{L}^{\nu ,\nu ^{\prime }}(\omega )=-2\mbox{Im}[\Sigma ^{R,\nu
\nu ^{\prime }}(\omega )_{1,1}]$ and $\Gamma _{R}^{\nu ,\nu ^{\prime
}}(\omega )=-2\mbox{Im}[\Sigma ^{R,\nu \nu ^{\prime }}(\omega )_{N,N}]$ with 
$\nu ,\nu ^{\prime }=g,f$. The Fermi functions $f_{\alpha }(\omega )$, with $%
\alpha =L,R$ depend on the temperatures $T_{L}$ and $T_{R}$ of the left and
right chains respectively: $f_{\alpha }(\omega )=1/(1+e^{\omega /T_{\alpha
}})$, in units where $k_B=1$. 

Finally, the lesser counterparts of Eqs. (\ref{con2}) and (\ref{con3}),
which correspond to Green's functions with one of the coordinates in the
central ($C$) chain and the other one in the left ($L$) or right ($R$)
chain, can be calculated by recourse again to Langreth rules \cite%
{langr,hern} 
\begin{eqnarray}
G_{L1,Cl^{\prime }}^{<} &=&G_{L,1,1}^{0,<}w^{\prime
}G_{C1,Cl^{\prime }}^{A}  +G_{L,1,1}^{0,R}w^{\prime }G_{C1,Cl^{\prime }}^{<} 
\nonumber \\
&&+\overline{F}_{L,1,1}^{0,<}w^{\prime }F_{C1,Cl^{\prime
}}^{A}  +\overline{F}_{L,1,1}^{0,R}w^{\prime }F_{C1,Cl^{\prime
}}^{<},  \label{conles2} \\
G_{R1,Cl^{\prime }}^{<} &=&g_{R,1,1}^{0,<}w^{\prime
}G_{CN,Cl^{\prime }}^{A} +g_{R,1,1}^{0,R}w^{\prime }G_{CN,Cl^{\prime }}^{<}.
\label{conles3}
\end{eqnarray}

\subsection{Heat currents and transmission functions}
We focus on the expression for the heat current evaluated in the contact
between the central chain $C$ and $R$ given in Eq. (\ref{jq2}). Using
 Eq. (\ref{conles3}) one obtains
\begin{eqnarray}
J^Q &=&-2 |w' |^2 \int_{-\infty }^{+\infty }\frac{d\omega }{2\pi }\omega \mbox{Re}%
[G_{CN,CN}^{<}(\omega )g_{R,1,1}^{0,A}(\omega )  \nonumber \\
&&+G_{CN,CN}^{R}(\omega )g_{R,1,1}^{0,<}(\omega )].  \label{jal2}
\end{eqnarray}%
Using (\ref{gcles}) and after some algebra (see Ref. \onlinecite{lilisup}),
it is found 
\begin{eqnarray}
J^Q &=&\int_{-\infty }^{+\infty }\frac{d\omega }{2\pi }\omega \lbrack
f_{L}(\omega )-f_{R}(\omega )][T^{n}(\omega )-T^{a}(\omega )],  \nonumber \\
T^{n}(\omega ) &=&\Gamma _{R}^{gg}(\omega )|G_{C,N,1}^{R}(\omega
)|^{2}\Gamma _{L, {\mbox{\it eff}}}^{gg}(\omega ),  \nonumber \\
T^{a}(\omega ) &=&\Gamma _{R}^{gg}(\omega )
| \overline{\Lambda }_{N,N}(\omega )|^{2} \Gamma _{R}^{ff}(\omega ),
\label{jqal3}
\end{eqnarray}%
where 
\begin{eqnarray}
\!\!\!\!\!\!&& \Gamma_{L,\mbox{{\it eff}}}^{gg} =\Gamma _{L}^{gg}+2\mbox{Re}%
[\Gamma _{L}^{gf}\Lambda _{1,N}^{A}]  
+|\Lambda _{1,1}^{R}|^{2}\Gamma _{L}^{ff}+|\Lambda
_{1,N}^{R}|^{2}\Gamma _{R}^{ff},  \nonumber \\
&&\overline{\Lambda }^R_{N,N} =G_{C,N,1}^{R}\Lambda^R_{1,N}.  \label{gamlam}
\end{eqnarray}

\begin{figure}
\centering   
\includegraphics
[width=70mm,height=90mm,angle=-90]{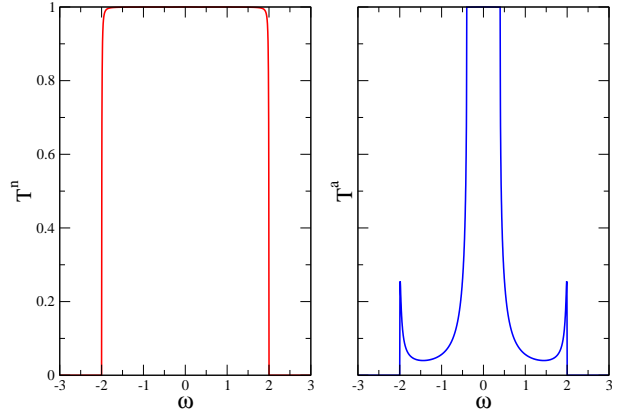} {\small {} }
\caption{{
(Color online)``normal'' (left)  and ``anomalous'' (right) 
transmission functions.
The latter is the counterpart to Andreev transmission function in 
electronic systems.
Parameters are $\Delta_L=0.2$, $J_L=J_R=J_C=J'=1$ and all chemical potentials set to zero.}}
\label{tt}
\end{figure}

The difference of Fermi functions in the expression of $J^Q$, reflects the fact that
 the existence of a 
non-vanishing heat current through the central system depends on the existence of
a difference of temperatures between the left and right chains.
The details of the model are enclosed in the behavior of the ``normal'' and
``anomalous''
 transmission functions $T^{n}(\omega )$ and $T^{a}(\omega )$, which  are analogous to
those defined in Ref. \onlinecite{lilisup} in the context of particle
transport in a setup with normal and superconducting wires. 
The first function has, in fact, the structure of a transmission. Notice that
it depends on the densities of states of the right and left chains through the
functions $\Gamma_R$ and $\Gamma_{L}^{eff}$, and one the particle propagator between the
first and last points of the central chain.
Instead, $T^{a}(\omega )$ actually
has the structure of a reflection process. Notice that it depends on the density of
states for particles and holes of the right reservoir and on a multiparticle
propagator $\overline{\Lambda }^R_{N,N}$ at the last point of the central chain.
Typical plots
for these functions are shown in Fig. \ref{tt}. These functions do not depend on
the temperatures $T_L$ and $T_R$ and are non-vanishing
only within a finite range of energies of a width that is set by the largest exchange
parameter between the left, right and central chains.
These functions are symmetric with respect
to $\omega=0$ for $B_L=B_R=0$ 
(see Fig. \ref{tt}). 
This symmetry is broken for finite
$B_{\alpha}$, since
the effect of a finite magnetic field in one of the side chains
is to shift the corresponding function as $\Gamma^{\nu \nu'}_{\alpha}(\omega) \rightarrow
\Gamma^{\nu \nu'}_{\alpha}(\omega-B_{\alpha})$. 

In the language 
of fermionic systems, two different kinds of processes take place in a 
normal-superconductor junction.
For energies higher than the gap, the transport is due to the tunneling of normal single particle
high-energy excitations. This mechanism contributes to the electronic transmission function
 $T^n(\omega)$. Instead, for low energies, below the gap, the transport is due to
the mechanism known as  ``Andreev reflection'', which implies 
the combination
of two fermions of the normal side into a Cooper pair within the
superconducting one, leaving a hole that is reflected  back from the junction
into the normal side. Because of this mechanism, 
$T^n(\omega) \sim T^a(\omega) \sim 1 $ for energies within the
superconducting gap, i.e. $|\omega| \leq \Delta_L$. 
The effective conversion of electrons into Cooper pairs
taking place in the mechanism of Andreev reflection helps to partice transport.
Mathematically, this is reflected by the fact that the total particle
transmission function is $T^n(\omega)+T^a(\omega)$. \cite{lilisup,btk}
Instead, in the case of heat transport, $T^a(\omega)$ and $T^n(\omega)$ 
contribute with opposite sign, as explicitly shown in 
Eq. (\ref{jqal3}),
i.e. the mechanism of Andreev reflection, plays a negative role regarding the heat transport. The consequence
is a vanishing heat transport due to excitations
within the energy window defined by the superconducting
gap.

In the original language of interacting spins, the above picture translates
as follows. Low energy spin excitations traveling from the isotropic chain
via flip-flop processes  in the $z$ direction meet an energy gap at the
other side of junction due to the anisotropic interaction which tends to
favor flip-flop processes  in a different direction. This favors the
simultaneous raising or lowering of two spins at two neighboring positions of
the chain and causes multiscattering processes in which a portion of the incident 
spin wave
packet manages to twist and propagate into the other side, at the same time
that a portion becomes reflected and propagates back. 

We can describe the behavior of $J^Q$ for low $T$ and small temperature gradients $\delta T$ as follows. 
Writing $T_R=T$ and $T_L=T_R+\delta T$ we can approximate the difference of Fermi functions in Eq. (\ref{jqal3}) as 
\begin{equation}
f_L(\omega)-f_R(\omega)\sim \frac{\partial f_R(\omega)}{\partial T} \delta T .
\end{equation}
On the other hand, from Fig. \ref{tt}, we can write,

\begin{eqnarray}
T^n(\omega)-T^a(\omega) \sim &0 & \omega \leq 2\Delta_L \\
T^n(\omega)-T^a(\omega)\sim & 1 &2\Delta_L \leq \omega \leq 2J
\end{eqnarray}
leading to,
\begin{eqnarray}\label{expo}
J^{Q} &=& \frac{\delta T }{\pi} \int_{2\Delta_L}^{+2J}d\omega \omega 
\frac{\partial f_R(\omega)}{\partial T} 
\end{eqnarray}
For low enough temperatures, $T\ll2\Delta_L \ll 2J$, this expression can be further approximated as,
 \begin{eqnarray}\label{expo2}
J^{Q} &\sim& \frac{\delta T }{\pi} \frac{\partial }{\partial T}\int_{2\Delta_L}^{\infty}d\omega \omega e^{-\beta \omega}
\\ 
& \sim &  \frac{4}{\pi} \delta T \frac{
\Delta_L^2}{T} e^{-\frac{2\Delta_L}{T}}\;\;\;\; 
\end{eqnarray}

Therefore, for $T<\Delta_L$, the heat current is exponentially small.

On the other hand, for $\Delta_L= 0$,  the behavior of the $J^Q$
is fully due to normal tunneling. For low $T$ we 
can perform a Sommerfeld expansion on the Fermi function to get 
\begin{eqnarray}\label{linear}
J^{Q} &=&  \frac{2}{T^2} \delta T \frac{\partial }{\partial \beta} (
\int_{0}^{+\infty }\frac{d\omega }{2\pi }\omega  f(\omega ) T^{n}(\omega ) )\nonumber \\
& \sim & \frac{ \pi}{3} T \delta T , \;\;\;\; T \ll J_{\alpha},\;\; \Delta_L = 0.
\end{eqnarray}

\section{Results}
In this section we discuss the behavior of the heat current as a function of the different 
ingredients of the spin system. For simplicity, we consider identical exchange parameters
along the left,
central and right chains: $J_L=J_C=J_R =J'=J$. Without loss of generality we set $J=1$. Thus, we 
focus on 
a spin heterostructure
with a single junction between a semi-infinite XX and a 
semi-infinite 
XY chain, which
in the fermionic language translates to a single S-N junction. For this particular 
configuration, our results do not depend on the length of the central chain. 

\begin{figure}
\centering   
\includegraphics
[width=70mm,height=90mm,angle=0]{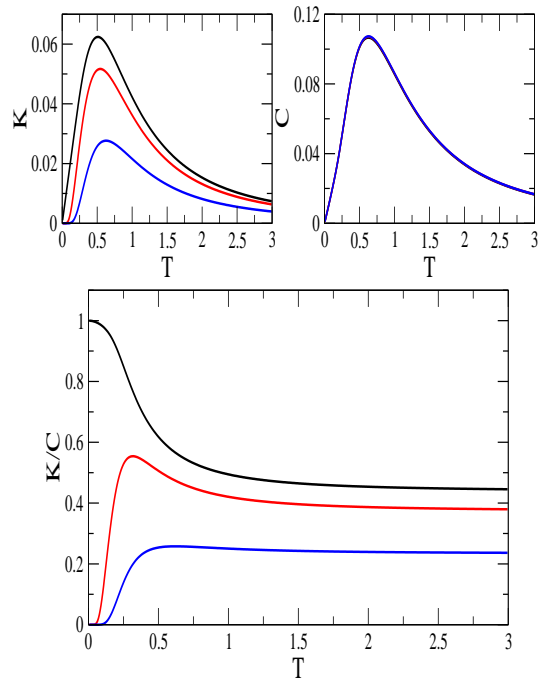} {\small {} }
\caption{{ 
(Color online) Top left: Conductance, 
defined as the ratio of heat current to temperature
gradient between right and left chains. 
Top right: specific heat of the central system.
Bottom: ratio of the above magnitudes.
Parameters are $T_L=T+\protect\delta_T, T_R=T$ where $\protect\delta_T=0.005$, 
all chemical potentials set to zero, and 
(from top to bottom) 
$\Delta_L=0,0.25,0.5$. }}
\label{conductividad}
\end{figure}

\begin{figure}
\centering   
\includegraphics
[width=70mm,height=90mm,angle=-90]{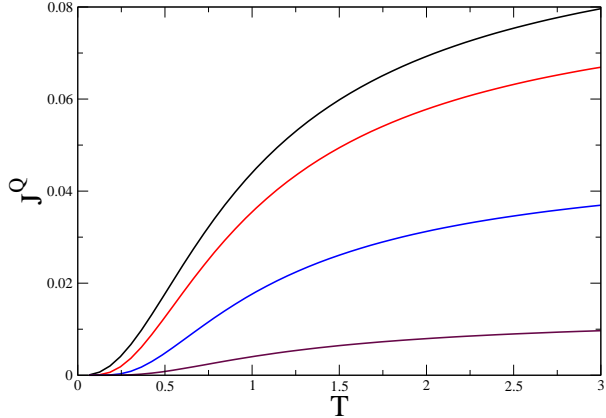}
\caption{{
(Color online) Heat current as a function of the temperature of the left-chain (right chain at fixed $T_R=0$) 
for all chemical potentials set to zero, all 
hopping  parameters set to 1 and different values of the paring parameter $\Delta_L$, 
from top to bottom $\Delta_L=0.1,0.25,0.5,0.75$. 
}}
\label{curr2}
\end{figure}

As discussed in the previous subsection, the
structure of the expression (\ref{jqal3}) for the heat current clearly
reflects the fact that for small temperature differences, we obtain 
a behavior of the form 
\begin{equation}
J^Q=-K \delta T,
\end{equation}
where the coefficient $K$ can be interpreted as a {\em thermal conductance}.
It is tempting to relate this coefficient with the 
conductivity $\kappa $
evaluated in
several works on the basis of linear response theory. 
If we assume that the relation between the two coefficients is similar to the
one between electrical conductance and electrical conductivity, $K$ and
$\kappa$ should differ just by a geometrical factor. However, to the best of our
knowledge, a rigorous relation between these two coefficients has not been
presented in the literature.
Nevertheless, the behavior of $K$ as a function of $T$ shown in the left upper panel of
Fig. \ref{conductividad}
for the case of two connected XX chains (see the plot corresponding to
$\Delta_L=0$) is similar to the one reported in the literature for
homogeneous and isotropic chains \cite{lista2}, \cite{lista5},\cite{lista3}. In this case, the 
anomalous component is zero and $K$
increases linearly in $T$ for low temperature 
[see Eq. (\ref{linear})], 
as discussed at the end of 
the previous section. The conductance reaches a maximum at $T \sim J$.
and decreases at higher temperatures, as a consequence of the finite bandwith
(energy window) for the spin excitations amenable to cross the central chain
transporting energy from one side to the other one. 
As expected, for a fixed
temperature $K $ decreases for increasing values of $\Delta_L $.
In agreement with the behavior discussed in the previous section, 
$K$ is exponentially small for $T<\Delta_L $  
[see Eq. (\ref{expo})], 
For higher temperatures,
the high energy excitations are allowed to perform 
tunneling above the energy gap, with the concomitant increase of $K$.
As in the case with $\Delta_L=0$, the maximum is achieved at $T\sim J$.  

We also evaluate the specific heat for the {\em equilibrium} central system
in contact to the side chains at the
same temperature $T$ as follows:

\begin{equation}
C(T) = - \frac{2}{N} \sum_{l=1}^N \int_{-\infty}^{+\infty}\frac{d \omega}{2 \pi}
\frac{\partial  f(\omega)}{\partial T} \omega
\mbox{Im}[G^R_{C,l,l}(\omega) ]
\end{equation}

In a normal metallic system as described by Drude model, this quantity
is related to the thermal conductivity 
through
$\kappa = v l C/3$, being $v$ the Fermi velocity of the electrons and
$l$ their mean free path \cite{aschcroft-mermin}. We plot this quantity in the right upper panel of Fig. 
\ref{conductividad}. This physical quantity is almost insensitive to the opening of the energy gap  and the
different plots, corresponding to different values of $\Delta_L$ almost coincide within the scale of the figure.
From the lower panel of Fig. \ref{conductividad} we see that while for  high temperatures there is a 
linear relation between $K$ and $C$, this is not the case at lower temperatures where Andreev type processes are relevant.

As stressed before, our calculation is not restricted to small temperature gradients.
We show in Fig. \ref{curr2} a plot of the heat current for several values of 
the anisotropy
parameter $\Delta_L$ as a function of the temperature
of the left chain while the temperature of the right chain $T$ is set fixed to zero.
The figure clearly shows the suppression of the current as a consequence of 
the
\textquotedblleft
Andreev reflection
\textquotedblright
 phenomena mentioned before.
In fact for $T < \Delta_L$ the current is exponentially small, while it grows 
for higher temperatures.

\begin{figure}
\centering   
\includegraphics
[width=70mm,height=90mm,angle=-90]{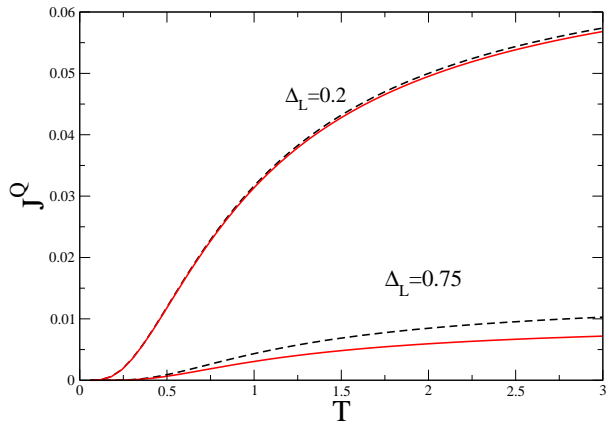} 
\caption{{(Color online) Thermal Diode effect. We show the heat current 
when ($\mu_L=0.3, T_L=T, \mu_R=0, T_R=0$) (dotted line) and when ($\mu_R=0.3, T_R=T, \mu_L=0, T_L=0$) 
for two values of the pairing parameter $\Delta_L=0.2,0.75$}}
\label{asy3}
\end{figure}

Finally, in Fig. \ref{asy3} we illustrate the behavior of the heat current
when finite different magnetic fields are applied at the two side chains.
The effect of applying magnetic fields at both sides of the junction
leads to a interesting effect which me name ``thermal diode effect''. 
As discussed in the previous section, a finite magnetic field originates
a shift in the arguments of the  functions $\Gamma_{\alpha}^{\nu \nu'}(\omega)$,
which 
leads to 
asymmetries in the transmission functions $T^n(\omega)$
and $T^a(\omega)$. For $\Delta_L=0$, only the normal transmission function
and the functions $\Gamma^{gg}_{\alpha}(\omega)$ are non-vanishing.
Furthermore, these functions are identical and gapless for $\alpha=L,R$.
Therefore, the heat flow is perfectly antisymmetrical (the sign of the current
is reversed preserving the absolute value) under the simultaneous
change $T_L \leftrightarrow T_R$ and $B_L \rightarrow B_R$.
Instead, for a finite $\Delta_L$, the situation changes. A gap opens for
the excitations of the left chain and the functions $ \Gamma^{\nu \nu'}_{L}(\omega) =0$
vanish for $|\omega| < \Delta_L $, while the functions
$\Gamma^{g g}_R(\omega)=\Gamma^{ff}_R(-\omega)$ remain finite. The consequence is
an asymmetry in the behavior of the transmission functions under the change
$B_L \rightarrow B_R$. The result is an effect of thermal rectification.
That is, the magnitude of  the current $J^Q$ 
when ($T_L=T$,$\mu_L=\mu$) and ($T_R=T'$,$\mu_R=\mu'$) is different to  $J^{Q'}$ 
when ($T_L=T'$,$\mu_L=\mu'$) and ($T_R=T$,$\mu_R=\mu $), which means that
the device 
is more likely 
to conduct heat when the temperature difference
is applied in one direction than in the other. We display the phenomena for two 
different values of $\Delta_L$. We 
show the current when ($\mu_L=0.3, T_L=T, \mu_R=0, T_R=0$) 
with dots and the current when ($\mu_R=0.3, T_R=T, \mu_L=0, T_L=0$) with a full line. 
When the value of $\Delta_L=0.2$ both currents are rather large and similar but when
 $\Delta_L=0.75$ the currents are smaller and clearly different.

\section{Summary and discussion}
We have presented a theoretical framework to study heat transport in one-dimensional spin 
heterostructures.

In the present work we have focussed on a simple system composed of  
a junction 
between an anisotropic (XY) and an isotropic (XX) chain
under the effect of an inhomogeneous magnetic field along the $z$ direction. Using the  
Jordan-Wigner 
transformation to map the problem into a fermionic system and using the non equilibrium  
Keldysh-Schwinger 
formalism we have obtained exact expression for the heat current in terms 
of Green's functions of the "disconnected" spin chain components. The resulting expressions  
can be evaluated numerically in a simple way. 
In the limits $\Delta_L \ll T \ll J$ and 
$T \ll \Delta_L \ll J$ explicit analytic expressions can also be given. 
We have studied the heat transport as a function of the  
different 
parameters of the model and we have shown that when  
different 
magnetic fields are applied at the end chains, 
a rectifying effect in the
heat current occurs. This effect might be of interest for applications.
Its origin can be traced back to the appearance of paring terms induced by the  
anisotropy parameter, 
which are 
in turn responsible 
for an Andreev 
reflection type mechanism.

In this work we have analyzed a simple model. However this methodology can be straightforwardly extended to more complex structures
with many junctions and disorder. Our treatment relies on the Jordan-Wigner transformation
which maps the original spin Hamiltonians into fermionic ones. In the case we have 
considered,
the latter are bilinear. In more generic models, although we expect the rectifying effect still 
to be 
present, the technical analysis could be more complicated.
For instance,
in the isotropic Heisenberg model, which in addition to the exchange interaction
along $x$ and $y$ directions, contains an additional exchange term along the $z$ direction, 
the Jordan-Wigner transformation translates such a term into a many-body fermionic
interaction, which does not enable a straightforward analytical solution of the problem,
as in the case we considered here. Nevertheless, the Green's function formalism offers
a framework for the construction of systematic approximations to treat those terms. 
Numerical methods could also be useful to deal with models containing many-body terms. \cite{feiguin}
As in electronic systems,  
many-body 
terms are expected to introduce further inelastic
scattering processes, which could add further ingredients in addition to the transport
mechanisms we have discussed here. We hope to report on some of  
these 
issues in future work.

\section*{Acknowledgments}

This investigation was sponsored by PIP 5254 and PIP 112-200801-00466  of CONICET, PICT 2006/483 of
the ANPCyT and projects X123 and X403 from UBA. 
We are partially supported by CONICET. G.S.L and L.A thanks C.Batista, D.Cabra, and D.Karevski for interesting comments.

\appendix

\section{Green's functions for an open chain with $p$-wave superconductivity}

In this appendix we show a derivation of the Green's functions $G_{L,1,1}^{0}(\omega )$, $\overline{G}_{L,1,1}^{0}(\omega )$, $%
F_{L,1,1}^{0}(\omega )$ and $\overline{F}_{L,1,1}^{0}(\omega )$, 
entering Eq. (\ref{sigmas}), 
which correspond to the end of a half infinite chain
with $p$-wave superconductivity in the BCS approximation.
  Making the
superconducting parameter $\Delta_L =0$, the first two Green's 
functions give the corresponding result for the normal chain 
$g_{R,1,1}^{0}(\omega )$ and  
$\overline{g}_{R,1,1}^{0}(\omega )$ respectively.

The Green's functions of the open chain can be solved considering a ring of $N$ sites,
periodic except for the fact that the energy at one site (which we label as
site 0) is increased by an energy $A$, and then taking the limit $%
N,A\rightarrow +\infty $. The Hamiltonian is

\begin{eqnarray}
H &=&\sum_{l=0}^{N-1}\{w[f_{l}^{\dagger }f_{l+1}+f_{l+1}^{\dagger }f_{l}] 
\nonumber \\
&&+\Delta \lbrack f_{l}^{\dagger }f_{l+1}^{\dagger }+f_{l+1}f_{l}]\} 
\nonumber \\
&&-\mu \sum_{l=0}^{N-1}f_{l}^{\dagger }f_{l}+Af_{0}^{\dagger }f_{0}.
\label{ah}
\end{eqnarray}

We have solved the problem using two different methods: i) solving the
equations of motion in Fourier space, and ii) solving a Dyson's equation that relates the above Green's
functions to those of the periodic chain ($A=0$) which can be obtained
easily using Bloch theorem. Both results of course coincide, but
the latter method involves a simpler algebra. We define a matrix 
\begin{equation}
 \tilde{G}  =
\left(
\begin{array}{cc}
G_{L,1,1}^{0}(\omega ) & F_{L,1,1}^{0}(\omega )\\
 \overline{F}_{L,1,1}^{0}(\omega ) &\overline{G}_{L,1,1}^{0}(\omega ) 
\end{array} 
\right),
\end{equation}
with the Green's functions 
for $A\neq 0$, and a corresponding matrix
$\tilde{g}$ for $A = 0$. These matrices are equivalent to the ones
obtained by using Nambu's representation for the Hamiltonian and the 
Green's functions. \cite{cuevas}
From the equations of motion
of these Green's functions, one obtains

\begin{equation}
\tilde{G}=\tilde{g}+\tilde{g}\tilde{A}\tilde{G}\text{,}  \label{adys}
\end{equation}%
where $\tilde{A}$, is proportional to $A$. 

 Solving Eq. (\ref{adys}) for $\tilde{G}$  and taking the limit $A \rightarrow +\infty$
the following expressions result

\begin{eqnarray}
G_{L,1,1}^{0} &=&h_0(\omega )-\frac{h_1^{2}(\omega )}{%
h_0(\omega )}-\frac{h_2^{2}(\omega )}{h_0^{\ast }(-\omega )%
}.  \label{aaa} \\
F_{L,1,1}^{0}&=&h_2(\omega )\left( \frac{h_1^{\ast }(-\omega )}
{h_0^{\ast }(-\omega )}-\frac{h_1(\omega )}{h_0(\omega )}
\right) .  \label{aba}
\end{eqnarray}
\begin{equation}
\overline{G}_{L,1,1}^{0}(\omega)=-(G_{L,1,1}^{0})^*(-\omega)
\,\,\,
\overline{F}%
_{L,1,1}^{0}=-(F_{L,1,1}^{0})^*(-\omega)
\end{equation}

The $h$-functions entering the second members of Eqs. (\ref{aaa}) and (\ref{aba}) are Green's functions of the periodic chain and can be calculated easily in Fourier space. The result is

\begin{eqnarray}
h_0(\omega ) &=&\frac{1}{N} \sum_{k} \frac{\omega +\epsilon _{k}}{\omega ^{2}-\epsilon _{k}^{2}-\Delta _{k}^{2}},  \nonumber \\
h_1(\omega ) &=&\frac{1}{N} \sum_{k} \frac{(\omega +\epsilon
_{k})\cos k}{\omega ^{2}-\epsilon _{k}^{2}-\Delta _{k}^{2}},  \nonumber \\
h_2(\omega ) &=&\frac{1}{N} \sum_{k} \frac{2\Delta \sin ^{2}k}
{\omega ^{2}-\epsilon _{k}^{2}-\Delta _{k}^{2}},  \label{agp}
\end{eqnarray}
where in the second members $\omega $ includes an infinitesimally small
imaginary part, $\epsilon _{k}=2w\cos k-\mu $, and $\Delta _{k}=2\Delta \sin
k$.

For $N\rightarrow +\infty $, the sums can be replaced by integrals.
Decomposing the integrands into a sum of simple fractions with denominators
linear in $\cos k$ and numerators independent of $k$, the integrals can be
evaluated analytically using \cite{gra}

\begin{equation}
I(b)=\frac{1}{\pi } \int_{0}^{\pi } \frac{dk}{\cos k+b}=\frac{1}{\sqrt{%
b^{2}-1}},  \label{ain}
\end{equation}%
where the sign of the root is determined by the sign of the imaginary part
of the second member.

Defining

\begin{eqnarray}
\tilde{\omega} &=&\frac{\omega }{2w},\text{ }\tilde{\Delta}=\frac{\Delta }{w},
\text{ }\tilde{\mu}=\frac{\mu }{2w},\text{ }d=1-\tilde{\Delta}^{2}, 
\nonumber \\
r &=&\left[ (\tilde{\omega}^{2}-\tilde{\Delta}^{2})d+(\tilde{\Delta}
\tilde{\mu})^{2}\right] ^{1/2}/d.  \nonumber \\
b_{1} &=&\tilde{\mu}/d-r,\text{ }b_{2}=\tilde{\mu}/d+r,  \nonumber \\
d_1 &=&\left[ 4wrd\right] ^{-1},  \label{aux}
\end{eqnarray}
the result takes the form

\begin{eqnarray}
h_0(\omega ) &=&d_1 [(r-\tilde{\omega}-\tilde{\mu}\tilde{\Delta}^{2}/b)I(b_{1})  \nonumber \\
&&+(r+\tilde{\omega}+\tilde{\mu}\tilde{\Delta}^{2}/b)I(b_{2})],  \nonumber \\
h_1(\omega ) &=&-d_1 \{2r-[(r-\tilde{\omega}-\tilde{\mu}\tilde{\Delta}^{2}/b)b_{1}I(b_{1}) 
\nonumber \\
&&+(r+\tilde{\omega}+\tilde{\mu}\tilde{\Delta}^{2}/b)b_{2}I(b_{2})]\}, 
\nonumber \\
h_2(\omega ) &=&d_1 \tilde{\Delta}\{2r+[I(b_{1})]^{-1}-[I(b_{2})]^{-1}\}.  \label{agp2}
\end{eqnarray}


\end{document}